\documentclass[12pt,twoside]{article}
\usepackage{epsfig}
\usepackage{graphicx}
\usepackage{caption}
\usepackage{subcaption}
\usepackage{amssymb}
\usepackage{url}
\usepackage{amsfonts,amsbsy}
\usepackage{latexsym}
\usepackage{amsmath}
\usepackage{subfig}
\usepackage{xcolor} 

\title{Plane-Wave Propagation\\ in Extreme Magnetoelectric (EME) Media}
\author{I.V. Lindell,{${}^1$} A. Sihvola{${}^1$} and A. Favaro{${}^2$}} 
\date{{${}^1$}Department of Radio Science and Engineering,\\ Aalto University, Espoo, Finland\\ {${}^2$}The Blackett Laboratory, Department of Physics,\\ Imperial College London, London, UK\\
{\tt ismo.lindell@aalto.fi}\\\vspace{-1pt}{\tt ari.sihvola@aalto.fi}\\\vspace{2.4pt} {\tt a.favaro@imperial.ac.uk}}
\parskip=\medskipamount
\def\e{\begin{equation}} 
\def\f{\end{equation}} 
\def\ea{\begin{eqnarray}} 
\def\fa{\end{eqnarray}} 

\def\##1{{\mbox{\textbf{#1}}}}
\def\%#1{{\mbox{\boldmath $#1$}}}
\def\=#1{{\overline{\overline{\mathsf #1}}}}

\def\*{^{\displaystyle*}}
\def\xx{\displaystyle{{}^\times}\llap{${}_\times$}}
\def\.{\cdot}
\def\x{\times}
\def\oo{\infty}

\def\ra{\rightarrow}

\def\Ra{\Rightarrow}

\def\l#1{\label{eq:#1}}
\def\r#1{(\ref{eq:#1})}
\def\am{\left(\begin{array}{c}}
\def\amm{\left(\begin{array}{cc}}
\def\ammm{\left(\begin{array}{ccc}}
\def\ammmm{\left(\begin{array}{cccc}}
\def\a{\end{array}\right)}

\def\A{\alpha}
\def\B{\beta}
\def\de{\delta}

\def\E{\epsilon}

\def\h{\eta}

\def\M{\mu}
\def\o{\omega}

\def\TH{\theta}

\def\VF{\varphi}

\def\tr{{\rm tr }}

\def\det{{\rm det}}

\begin{document}

\maketitle

\begin{abstract}
The extreme magnetoelectric medium (EME medium) is defined in terms of two medium dyadics, $\=\A$, producing electric polarization by the magnetic field and $\=\B$, producing magnetic polarization by the electric field. Plane-wave propagation of time-harmonic fields of fixed finite frequency in the EME medium is studied. It is shown that (if $\omega\neq 0$) the dispersion equation has a cubic and homogeneous form, whence the wave vector $\#k$ is either zero or has arbitrary magnitude. In many cases there is no dispersion equation (``NDE medium") to restrict the wave vector in an EME medium. Attention is paid to the case where the two medium dyadics have the same set of eigenvectors. In such a case the $\#k$ vector is restricted to three eigenplanes defined by the medium dyadics. The emergence of such a result is demonstrated by considering a {more regular} medium, and taking the limit of zero permittivity and permeability. The special case of uniaxial EME medium is studied {in detail}. It is shown that an interface of a uniaxial EME medium appears as a DB boundary when the axis of the medium is normal to the interface. More in general, EME media display interesting wave effects that can potentially be realized through metasurface engineering.
\end{abstract}

\section{Introduction}

For the general bi-anisotropic medium, conditions between the four electromagnetic field vectors can be expressed as 
\ea \#D &=& \=\A\.\#B + \=\E'\.\#E, \l{DBE}\\
\#H &=& \=\M{}^{-1}\.\#B + \=\B\.\#E, \l{HBE}\fa
in terms of four Gibbsian dyadics $\=\A,\=\E',\=\M{}^{-1}$ and $\=\B$. 

The dyadics $\=\A$ and $\=\B$ in \r{DBE} and \r{HBE} have been called magnetoelectric parameters in the past because they represent coupling between electric polarization and magnetic field on one hand and magnetic polarization and electric field on the other hand. Couplings of this kind were speculated by Pierre Curie already in 1894 \cite{Curie}, while the term magnetoelectric was coined by P. Debye in 1926 \cite{Debye}. The book \cite{Landau} by Landau and Lifshitz gave a systematic analysis of the magnetoelectric effect, in particular, concerning optical activity and the effect of the motion of the medium. The magnetoelectric effect was first measured by Astrov \cite{Astrov} in 1960. Physical media with magnetoelectric properties are also considered in \cite{ODell}.

In the present paper, we consider the extreme example of a medium with magnetoelectric parameters, by assuming {(over a limited range of frequencies)}
\e \=\E'=0,\ \ \ \ \=\M{}^{-1}=0. \f
The conditions \r{DBE} and \r{HBE} are now reduced to the simpler form
\e \#D = \=\A\.\#B, \ \ \ \ \#H=\=\B\.\#E, \l{ME}\f
and thus the number of free parameters is 9 ($\=\A$) + 9 ($\=\B$) =18. {We exclude cases of pathological nature, and demand that both $\=\A$ and $\=\B$ are invertible.} Since the magnetoelectric dyadics define the whole medium, it will be called by the name extreme magnetoelectric medium, or EME medium for short. Interest in ``extreme" cases of medium and boundary conditions has been created by the recent progress in metamaterial and metaboundary engineering \cite{Yaghjian2010,Istanbul2011,Alu2011,Radi2012}. Applications of metaboundaries (metasurfaces) include flat lenses, laser-beam deflectors and modulators, broadband waveplates, thin-film displays and holograms, as well as near-perfect absorbers, see the reviews \cite{Kildishev, Yu}. 

Media defined by \r{ME} are interesting because they generalize two medium classes which have been recently studied. The PEMC (perfect electromagnetic conductor) medium \cite{PEMC} can be defined as
\e \#D = M\#B,\ \ \ \ \#H=-M\#E, \l{PEMC}\f
or by choosing $\=\A=-\=\B=M\=I$ in \r{ME}. This medium is a generalization of the PMC (perfect magnetic conductor), defined by $M=0$, and the PEC (perfect electric conductor), defined by $1/M=0$. The PEMC is also known by the name axion medium \cite{Hehl,Hehl2008}. An interface between vacuum and PEMC is known to serve as a polarization rotator for the reflected fields \cite{PEMC}. Metaboundary realizations of the PEMC have been investigated \cite{Caloz,Elmaghrabi}. As another special case of the EME medium, the simple skewon medium \cite {SS} defined by
\e \#D = N\#B,\ \ \ \ \#H=N\#E, \l{SS}\f
corresponds to the choice $\=\A=\=\B=N\=I$. As was shown in \cite{SS}, the interface of such a medium can serve as a DB boundary, so that the normal components of $\#D$ and $\#B$ at the interface are zero \cite{DB}. The DB boundary can be used to achieve invisibility to the monostatic radar, whose transmitter and receiver are located at the same point. More in detail, by engineering an appropriate material coating, it is possible to reduce the backscattering cross-section of any object that is endowed with certain symmetries \cite{Sihvola2009, Lindell2009a}. In a similar context, we recall that the inner shell of the electromagnetic invisibility cloak is a DB boundary \cite{Zhang,Yaghjian08}. Metamaterial realizations of the DB boundary have been studied both theoretically and experimentally \cite{SS,Zaluski11,Zaluski14}. 

In the present article, we consider the effect of the medium dyadics $\=\A$ and $\=\B$, assumed constant, on time-harmonic plane-wave electromagnetic fields. The actual implementation of generic EME media is left as a subject for future work. One can however expect that, as metaboundaries are simpler to manufacture than bulk metamaterials \cite{Kildishev,Yu}, the early applications of EME media will be in planar and interface photonics.

According to Hehl and Obukhov, the medium dyadics of the EME medium can be decomposed in three parts as \cite{Hehl,Hehl2008,MDEM} 
\ea \=\A &=& \=\A_1+ \=\A_2+ \=\A_3, \\
\=\B &=& \=\B_1 + \=\B_2 + \=\B_3, \fa
respectively labeled as principal ($\=\A_1,\=\B_1$), skewon ($\=\A_2,\=\B_2$) and axion  ($\=\A_3,\=\B_3$) components. They can be defined by the conditions (see Appendix 1)
\ea \=\A_1 = -\=\B{}_1^T &=& \frac{1}{2}(\=\A-\=\B{}^T) - \frac{1}{6}\tr(\=\A-\=\B{}^T)\=I, \l{A1}\\
\=\A_2=\=\B{}_2^T &=& \frac{1}{2}(\=\A+\=\B{}^T), \l{A2}\\
\=\A_3=-\=\B_3 &=& \frac{1}{6}\tr(\=\A-\=\B{}^T)\=I. \l{A3}\fa
One can verify that the PEMC medium \r{PEMC} has only an axion component, while the simple-skewon medium \r{SS} has only a skewon component. A simple principal EME medium can be defined, for example, in terms of the symmetric trace-free dyadic $\=\A=-\=\B=A(\=I-3\#u\#u)$,  where $\#u$ is a unit vector. As an aside, the Hehl-Obukhov decomposition is relativistically covariant \cite{Hehl}, whereas the definition \r{ME} of EME media is not.

\section{Dispersion equation}

For a plane wave with $\#E(\#r) =\#E\exp(-j\#k\.\#r)$ dependence, the Maxwell equations in a source-free region of space can be written as
\ea \#k\x\#E &=& \o\#B , \l{kxE}\\
\#k\x\#H &=& -\o\#D. \l{kxH}\fa
To find solutions, let us apply the EME-medium conditions \r{ME} and eliminate the field components $\#D, \#H$ and $\#B$, whence the remaining equation has the form 
\e \=D(\#k)\.\#E=0, \l{DkE}\f
as defined by the dispersion dyadic
\e \=D(\#k) = \#k\x\=\B + \=\A\x\#k. \l{Dk}\f
For a solution $\#E\not=0$, the dispersion equation 
\e \det\=D(\#k) = \frac{1}{6}\=D(\#k)\xx\=D(\#k):\=D(\#k) =0, \l{detD}\f
must be satisfied by $\#k$ \cite{Methods}. Substituting \r{Dk}, and applying $\det(\#k\x\=\B)=\det(\=\A\x\#k)=0$ together with other dyadic rules from \cite{Methods}, one can expand \r{detD} so as to achieve the dispersion equation for the wave vector $\#k$, as
\ea \det\=D(\#k) &=& (\#k\x\=\B)^{(2)}:(\=\A\x\#k)+ (\#k\x\=\B):(\=\A\x\#k)^{(2)} \\
&=& (\#k\#k\.\=\B{}^{(2)}):(\=\A\x\#k)+ (\#k\x\=\B):(\=\A{}^{(2)}\.\#k\#k) \\
&=& \#k\.(\=\A\x\#k)\.(\=\B{}^{(2)T}\.\#k) + (\#k\.\=\A{}^{(2)T})\.(\#k\x\=\B)\.\#k \\
&=& (\=\A{}^T\.\#k)\.\#k\x(\=\B{}^{(2)T}\.\#k) + (\=\A{}^{(2)}\.\#k)\.\#k\x(\=\B\.\#k) \\
&=&  (\=\A{}^T\.\#k)\.(\=\B{}^T\x(\#k\.\=\B{}^T))\.\#k + \#k\.((\=\A{}^T\.\#k)\x\=\A{}^T)\.(\=\B\.\#k) \l{disp1}\\
&=& -(\=\A{}^T\.\#k)\.(\=\B{}^T\x\#k + \#k\x\=\A{}^T)\.(\=\B\.\#k) \\
&=& (\=\B\.\#k)\.(\=\A\x\#k + \#k\x\=\B)\.(\=\A{}^T\.\#k)\\
&=& (\=\B\.\#k)\.\=D(\#k)\.(\=\A{}^T\.\#k)=0. \l{disp}\fa
In making the step to \r{disp1}, the dyadic rule
\e \=A\.(\#a\x\=A{}^T) = (\=A{}^{(2)}\.\#a)\x\=I, \l{AaAT}\f
has been applied. We can make the following observations concerning the dispersion equation \r{disp}:

\begin{itemize}
\item It appears remarkable that the dispersion equation for EME media has cubic form, see \r{disp}, while the general one for bi-anisotropic media is quartic. Actually, using 4D electrodynamics \cite{Hehl,MDEM}, where $\#k$ is replaced by the four-vector $(\#k,\o)$, the dispersion equation of EME media is found to take the quartic form $\o\det(\=D(\#k))=0$, whence another solution is possible, $\o=0$. Here we have tacitly assumed that $\o\not=0$.

\item By making use of $\#k=k\#u$, where $\#u$ is a unit vector, the dispersion equation can be formulated as
\e \det\=D(\#k) = k^3F(\#u)=0,\l{kfu}\f
with
\e F(\#u) = (\=\B\.\#u)\.\=D(\#u)\.(\=\A{}^T\.\#u). \l{fu}\f
For any choice of $\#u$, one solution of \r{kfu}, and thus \r{disp}, is $k=0$. It follows that the electromagnetic fields are constant in space ($\#k=0$), but not in time ($\o\neq0$). Provided $k$ is nonzero, \r{kfu} can be interpreted as $F(\#u)=F(\TH,\VF)=0$ by means of spherical angular coordinates. Real solutions to this equation typically describe a closed curve, or multiple closed curves, on the unit sphere. Then, because the magnitude of $\#k$ is unconstrained, the dispersion surfaces extend from the origin ($k\rightarrow0$) to infinity ($k\rightarrow +\infty$) with the curves on the unit sphere as cross sections. Another possible scenario is that $F(\TH,\VF)=0$ for all angles $\TH$ and $\VF$, whereby the dispersion equation becomes just an identity. The corresponding electromagnetic media were studied in \cite{MDEM,NDE}, and dubbed NDE (No Dispersion Equation) media.
\item The form of \r{disp1} reveals that $\det\=D(\#k)=0$ is satisfied whenever $\#k$ is a right eigenvector of $\=\B$ or a left eigenvector of $\=\A$. Thus, real-valued eigenvectors are tangential to the dispersion surfaces.
\item Let us introduce the dyadics $\=\A_+$ and $\=\A_-$ by
\e \=\A_\pm = \frac{1}{2}(\=\A \mp \=\B{}^T), \f
whence we have
\e \=\A = \=\A_+ + \=\A_-,\ \ \ \ \=\B = -\=\A{}^T_+ + \=\A{}^T_-. \f
Defining $\=\B_\pm=\mp\=\A{}^{\hspace{1pt}T}_\pm$, one can verify that $\=\A_+,\=\B_+$ and $\=\A_-,\=\B_-$ correspond to the respective principal-axion and skewon parts of the EME medium, see Appendix 1. The two parts can be separated in the dispersion equation \r{disp} as
\ea \det\=D(\#k) &=& -2\#k\.(\=\A{}^T_+\.\#k)\x(\=\A{}^{2T}_+\.\#k) + 2 \#k\.(\=\A_-^T\.\#k)\x((\=\A_-\.\=\A_+)^T\.\#k) \nonumber\\
&-& 2\tr\=\A_-\ \#k.(\=\A_+^T\.\#k)\x(\=\A_-^T\.\#k)=0. \l{disp+-} \fa
\item From \r{disp+-} it is seen that for a skewon EME medium ($\=\A_+=0$), the equation is satisfied for any $\#k$ and, thus, the medium belongs to the class of NDE media. Actually, it is known that any skewon medium serves as an example of an NDE medium \cite{MDEM,NDE}. 
\item The PEMC \r{PEMC}, and simple skewon \r{SS} media are other examples of NDE media. 
\item If one of the dyadics $\=\A$, $\=\B$, or $\=\A{}^T\.\=\B$ is a multiple of the unit dyadic, the dispersion equation is again satisfied by any $\#k$.  
\item Principal-axion EME media are defined by $\=\A_-=0$, or $\=\B=-\=\A{}^T$. In this case the dispersion dyadic $\=D(\#k)= (\=\A\x\#k)^T +\=\A\x\#k$ is symmetric and the dispersion equation \r{disp} is reduced to the simple form
\ea \frac{1}{2}(\=\B\.\#k)\.\=D(\#k)\.(\=\A{}^T\.\#k) &=& -(\=\A{}^T\.\#k)\.(\=\A\x\#k)\.(\=\A{}^T\.\#k) \nonumber\\
&=& \#k\.(\=\A{}^T\.\#k)\x(\=\A{}^{2T}\.\#k)=0, \l{kBB}\fa
which requires that the vectors in the triple product be linearly dependent. When $\#k$ is not an eigenvector of $\=\A{}^T$, i.e., when $\#k\x(\=\A{}^T\.\#k)\not=0$, we can expand
\e \=\A{}^{2T}\.\#k = c_1\#k + c_2\=\A{}^T\.\#k \l{B2c1c2}\f
in terms of some scalars $c_1,c_2$. Requiring \r{kBB}, and hence \r{B2c1c2}, to be valid for any $\#k$ leads to 
\e \=\A{}^{2T} = c_1\=I + c_2\=\A{}^T.\f
Thus, for a principal-axion EME medium to be an example of an NDE medium, the dyadic $\=\A$ must satisfy a second-order equation. 
\end{itemize}

\section{Plane-wave fields}

Assuming that $\#k$ is a solution of \r{disp}, to find the field vectors of a plane wave in an EME medium, we can start from the Gauss laws for electricity and magnetism which, alongside \r{ME}, imply
\ea \#k\.\#B&=&0, \\
\#k\.\#D&=&\#k\.\=\A\.\#B = (\=\A{}^T\.\#k)\.\#B=0. \fa
If $\#k$ is not an eigenvector of $\=\A{}^T$, we must have
\e \o\#B = C\#k\x(\=\A{}^T\.\#k),  \l{oB}\f
where $C$ is some scalar. Substituting \r{oB} into \r{kxE} as
\e \#k\x\#E-\o\#B = \#k\x(\#E - C\=\A{}^T\.\#k)=0, \f
the vector $\#E$ must be of the form
\e \#E = C(\=\A{}^T\.\#k + A\#k), \l{E}\f
where $A$ is another scalar. The other fields can now be found from \r{ME} as 
\ea \#H &=& \=\B\.\#E = C(\=\B\.\=\A{}^T\.\#k + A\=\B\.\#k), \l{H}\\
 \o\#D &=& \=\A\.\o\#B = C\=\A\.(\#k\x(\=\A{}^T\.\#k))= C(\=\A{}^{(2)}\.\#k)\x\#k, \l{oD}\fa
where we have again applied the rule \r{AaAT}. Obviously, the fields \r{oB}, \r{E}, \r{H} and \r{oD} satisfy \r{ME} and \r{kxE}. Requiring that the condition \r{kxH} be also satisfied leads to
\ea \#k\x\#H+\o\#D &=& C\#k\x(\=\B\.(\=\A{}^T\.\#k) + A\=\B\.\#k -\=\A{}^{(2)}\.\#k) \nonumber\\
&=&C(\#k\x\=\B\.(\=\A{}^T\.\#k) -\#k\x\=\A{}^{(2)}\.\#k)+ CA\#k\x(\=\B\.\#k) \nonumber\\
&=&C((\#k\x\=\B)\.(\=\A{}^T\.\#k) {-(\=\A\x(\#k\.\=\A))\.\#k})+ CA\#k\x(\=\B\.\#k) \nonumber\\
 &=&C\=D(\#k)\.(\=\A{}^T\.\#k) + CA\#k\x(\=\B\.\#k)=0 . \l{kxHoD}\fa
One may achieve the same result by noting that, because of \r{disp} and 
\e \#k\.\=D(\#k)\.(\=\A{}^T\.\#k) = \#k\.(\=\A\x\#k)\.(\=\A{}^T\.\#k)= (\=\A{}^T\.\#k)\x\#k\.(\=\A{}^T\.\#k) =0, \f
the vector $\=D(\#k)\.(\=\A{}^T\.\#k)$ is a multiple of $\#k\x(\=\B\.\#k)$. The parameter $A$ can be solved from \r{kxHoD} as
\e A =-\frac{\#a\.\=D(\#k)\.(\=\A{}^T\.\#k)}{\#a\.(\#k\x(\=\B\.\#k))}, \l{A}\f
where $\#a$ may be any vector satisfying $\#a\.(\#k\x\=\B\.\#k)\not=0$. As an example, for the skewon medium with $\=\B=\=\A{}^T$ we obtain $A=-\tr\=\A$.  Inserting \r{A} in \r{E} and \r{H} completes the determination of the plane-wave fields for the EME medium. When $\#k$ is an eigenvector of $\=\B$, the formula \r{A} breaks down. Nevertheless, $\#E$ and $\#H$ are still specified by \r{E} and \r{H} with $A$ being arbitrary. The fields $\#B$ and $\#D$ are given by \r{oB} and \r{oD} in all cases.

In the above analysis $\#k\not=0$ has been tacitly assumed. However, since $\#k=0$ is a valid solution to \r{disp}, for completeness, we may briefly consider the possible fields in this case. From \r{kxE} and \r{kxH} we have $\#B=\#D=0$ while $\#E$ and $\#H$ may be nonzero. In fact, from \r{Dk} and \r{DkE} we see that $\#E$ is not restricted by the medium while $\#H$ is obtained from \r{ME}. Creating a {``wave"} with $\#k=0$ in the EME medium may, however, be a problem, because for a wave incident on its interface the transverse $\#k$ component is continuous through the interface.

\subsection{Skewon EME media}

For the skewon EME-medium ($\=\B=\=\A{}^T$), we have $(\#k\x\=\B)^T= -\=\B{}^T\x\#k= -\=\A\x\#k$. The dispersion dyadic is now antisymmetric and can be expressed as
\e \=D(\#k)= \=\A\x\#k -(\=\A\x\#k)^T = \#a(\#k)\x\=I,\ \ \ \ \ \ \#a(\#k)= (\tr\=\A) \#k-\=\A{}^T\.\#k . \l{DkAk}\f
Because the determinant of an antisymmetric dyadic vanishes, the dispersion equation is satisfied for any $\#k$, a well-known fact \cite{Hehl, NDE}. The electric field vector is obtained from $\=D(\#k)\.\#E = \#a(\#k)\x\#E=0$ as
\e  \#E = C(\=\A{}^T\.\#k -\tr\=\A\ \#k), \l{Esk}\f
in terms of some scalar $C$. Comparing with \r{E}, we have that $A=-\tr\=\A$, as noted above. The other fields are obtained from
\ea \#H &=& C(\=\A^{2T}\.\#k - \tr\=\A\ \=\A{}^T\.\#k), \l{HSk}\\
\o\#B &=& C\#k\x\=\A{}^T\.\#k, \l{BSk}\\
\o\#D&=& -C\#k\x\=\A{}^{(2)}\.\#k = -C\#k\x(\=\A{}^{2T}\.\#k -\tr\=\A\ \=\A{}^T\.\#k). \l{DSk}\fa

\subsection{Principal-axion EME media}
Let us consider the special case of principal-axion EME media, restricted by the condition $\=\A_-=0$, or $\=\B=-\=\A{}^T$ in \r{ME}. In this case the dispersion dyadic $\=D(\#k)$ is symmetric. The dispersion equation \r{disp} is now reduced to the form of \r{kBB} and the plane-wave fields \r{oB}, \r{E}, \r{H} and \r{oD} become
\ea \#E &=& C(\=\A{}^T\.\#k+A\#k), \\
\#H &=& - C(\=\A{}^{2T}\.\#k+A\=\A{}^T\.\#k), \\
\o\#B &=&  C\#k\x(\=\A{}^T\.\#k),\\
\o\#D &=& - C\#k\x(\=\A{}^{(2)}\.\#k). \fa
Applying the expansion  
\e \=\A{}^{(2)} = \=\A{}^{2T}- \tr\=\A\ \=\A{}^T + \tr\=\A{}^{(2)}\ \=I, \f
valid for any dyadic $\=\A$ \cite{Methods}, and assuming $\#k\x(\=\A{}^T\.\#k)\not=0$, in which case the dispersion equation \r{kBB} implies \r{B2c1c2}, the formula \r{A} can be reduced to 
\e A =-\frac{\#a\.\#k\x(\=\A{}^{2T}+\=\A{}^{(2)})\.\#k }{\#a\.\#k\x(\=\A{}^T\.\#k)}= -\frac{\#a\.\#k\x(2\=\A{}^{2T}\.\#k-\tr\=\A\ \=\A{}^T\.\#k) }{\#a\.\#k\x(\=\A{}^T\.\#k)} = \tr\=\A-2c_2, \l{A11}\f
and the electromagnetic fields can be rewritten as
\ea \#E &=& C((\tr\=\A-2c_2)\#k+ \=\A{}^T\.\#k),  \l{E1}\\
\#H &=& - C(c_1\#k+(\tr\=\A-c_2)\=\A{}^T\.\#k), \l{H1}\\
\o\#B &=&  C\#k\x(\=\A{}^T\.\#k),\l{oB1}\\
\o\#D &=& - C(\tr\=\A-c_2)\#k\x(\=\A{}^T\.\#k). \l{oD1}\fa

When $\#k$ is not an eigenvector of $\=\A{}^T$, the field vectors $\#E$ and $\#H$ are parallel to the plane defined by $\#k$ and $\=\A{}^T\.\#k$, while the vectors $\#D$ and $\#B$ are orthogonal to it. It is easy to check that, provided $\#k$ fulfills \r{kBB}, the field expressions \r{E1} -- \r{oD1} satisfy the conditions \r{ME}, \r{kxE} and \r{kxH} for any scalars $C$, $c_1$ and $c_2$. The fields $\#B$ and $\#D$ of a plane wave are parallel in any principal-axion EME medium.

\section{EME media with commuting dyadics}
As an example, let us consider a restricted class of EME media by requiring that the two dyadics $\=\A$ and $\=\B{}^T$ are of full rank and commute, i.e.\ 
\e \=\A\.\=\B{}^T=\=\B{}^T\.\=\A. \l{Commute}\f
From \r{A1} - \r{A3} it is obvious that both skewon EME media and principal-axion EME media belong to this {category}.

Because commuting dyadics have the same set of eigenvectors, they can be expressed as
\e \=\A= \sum_{i=1}^{3}\, \A_i \#a_i'\#a_i,\ \ \ \ \=\B = \sum_{i=1}^{3}\, \B_i \#a_i\#a_i'. \l{ABsum}\f
The vectors $\#a_i$ are assumed to make a basis satisfying $\#a_1\x\#a_2\.\#a_3=1$ and the vectors $\#a_i'$ make the reciprocal basis with the properties \cite{Methods}
\e \#a_1'=\#a_2\x\#a_3,\ \ \ \#a_2'=\#a_3\x\#a_1,\ \ \ \#a_3'=\#a_1\x\#a_2, \l{a123}\f
\e \#a_i'\.\#a_j=\de_{ij},\ \ \ \ \#a_1'\.\#a_2'\x\#a_3'=1, \f
\e \=I = \sum_{i=1}^{3}\, \#a_i\#a_i' = \sum_{i=1}^{3}\,\#a_i'\#a_i{.} \f
The most general EME medium has $9 (\=\A)+ 9 (\=\B)=18$ parameters. For the present class of EME media the number of parameters is reduced to $9 (\=\A) + 3 (\B_i)=12$. There are no principal and axion parts if the parameters satisfy $\A_i=\B_i$, while there is no skewon part for $\A_i=-\B_i$. In both of these cases the number of parameters is further reduced to $9$.

\subsection{Dispersion equation}
Inserting \r{ABsum} and $\#k=\sum_{i}\, k_i\#a_i$, the dispersion dyadic \r{Dk} can be written in the form
\ea \=D(\#k) &=&  -(\A_1+\B_2)k_3\#a_1'\#a_2'+ (\A_1+\B_3)k_2\#a_1'\#a_3' +
(\A_2+\B_1)k_3\#a_2'\#a_1' \\
&& -(\A_2+\B_3)k_1\#a_2'\#a_3' -(\A_3+\B_1)k_2\#a_3'\#a_1'+ (\A_3+\B_2)k_1\#a_3'\#a_2'.\fa
For $\#E=\sum_{i}\, E_i\#a_i$, equation \r{DkE} can be expressed in component form as
\e \ammm 0 & (\A_1+\B_2)k_3 & -(\A_1+\B_3)k_2\\
-(\A_2+\B_1)k_3 & 0 & (\A_2+\B_3)k_1\\
(\A_3+\B_1)k_2 & -(\A_3+\B_2)k_1& 0\a\am E_1\\ E_2\\ E_3\a= \am 0\\ 0\\ 0\a. \l{DEAB}\f
Requiring $\#E\not=0$, the determinant of the coefficient matrix must vanish, leaving us with the dispersion equation
\e k_1k_2k_3 f(\=\A,\=\B{}^T)=0, \l{dispAB}\f
where the function depending on the medium dyadics is defined by 
\e f(\=\A,\=\B{}^T)=(\A_1+\B_2)(\A_2+\B_3)(\A_3+\B_1) - (\A_1+\B_3)(\A_2+\B_1)(\A_3+\B_2)= -f(\=\B{}^T,\=\A). \l{fAB}\f
Let us consider different possibilities:

\begin{figure}[t]
\centering
    \begin{subfigure}[c]{0.3\textwidth}
        \centering
        \includegraphics[width=0.953\textwidth]{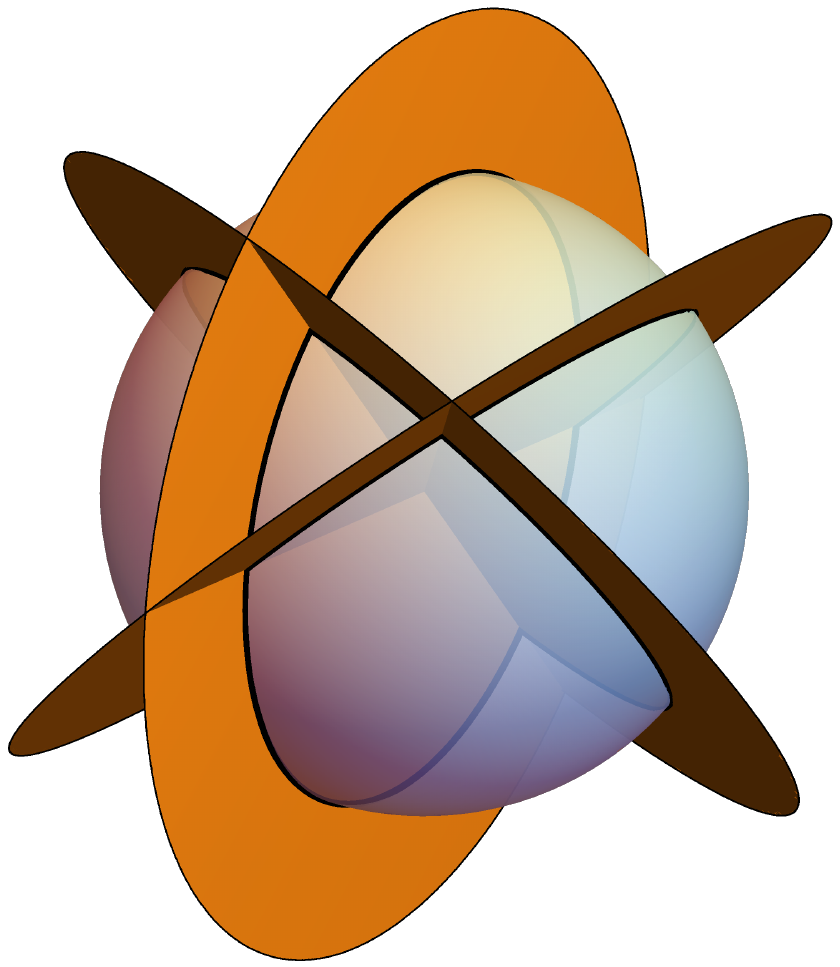}
    \end{subfigure}
    \qquad\qquad
    \begin{subfigure}[c]{0.3\textwidth}
        \includegraphics[width=0.952\textwidth]{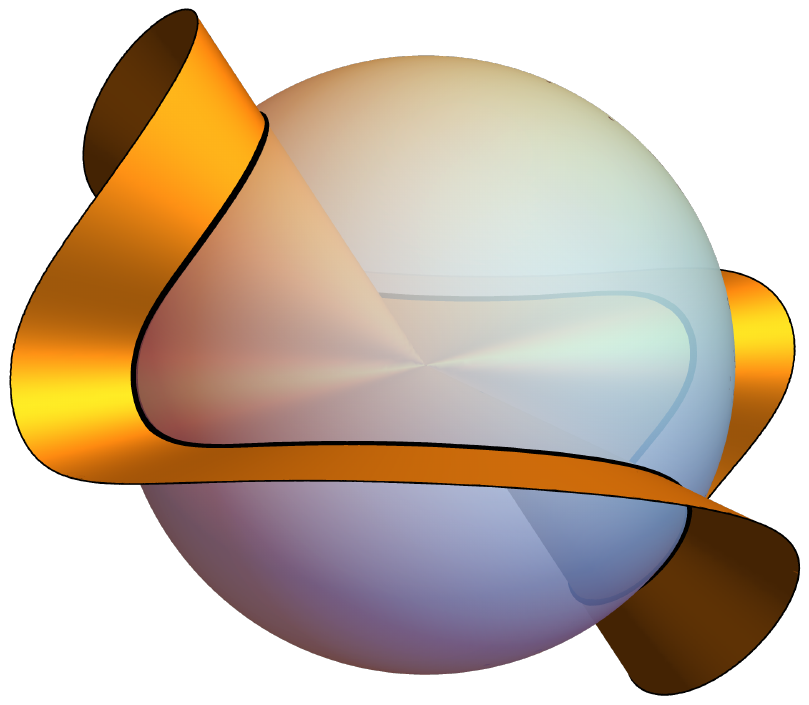}
    \end{subfigure}\\[5pt]
    \begin{subfigure}[c]{0.3\textwidth}
        \centering
        \includegraphics[width=0.75\textwidth]{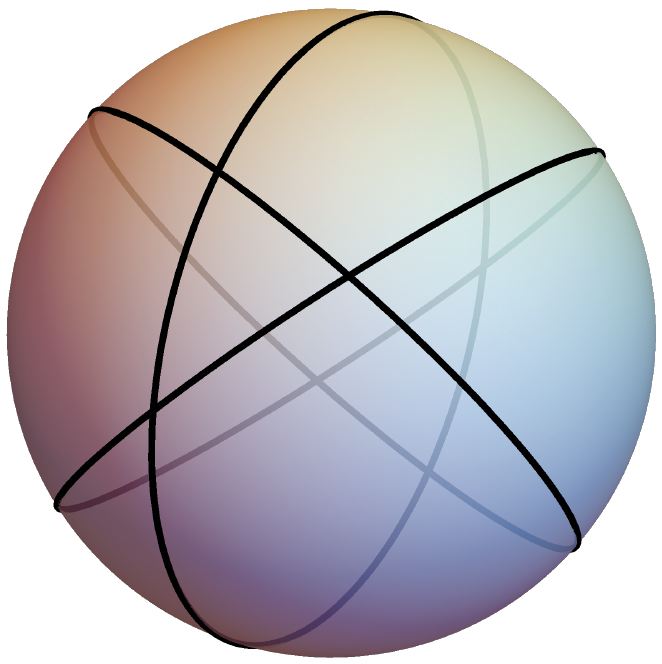}
        \caption{Commuting dyadics}
    \end{subfigure}
    \qquad\qquad
     \begin{subfigure}[c]{0.3\textwidth}
        \centering
        \includegraphics[width=0.747\textwidth]{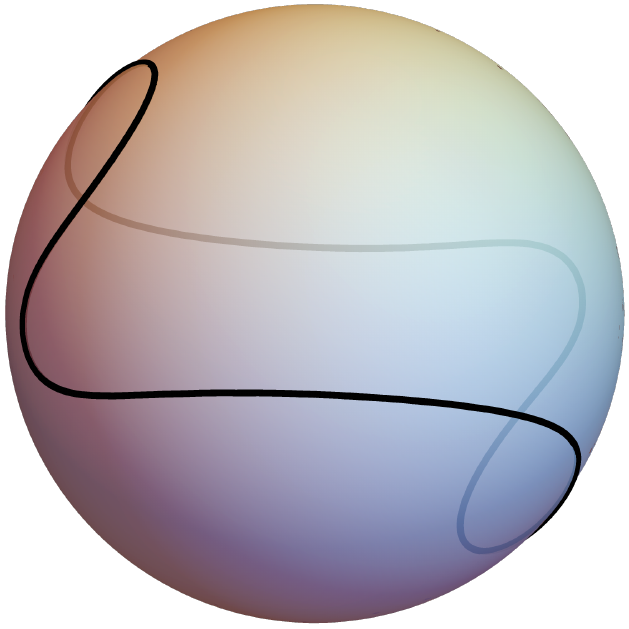}
        \caption{Randomly chosen dyadics}
    \end{subfigure}
    \caption{{Dispersion surfaces of EME media are described by closed curves on the unit sphere, and extend radially to infinity. We illustrate this for EME media such that (a) the dyadics $\overline{\overline{\A}}$ and $\overline{\overline{\B}}{}^{T}$ have coinciding eigenvectors and (b) the dyadics are chosen randomly. The intersections of circles on the unit sphere in (a) match the three eigenvector directions.}}
\end{figure}

\begin{figure}[t] 
    \centering
    \begin{subfigure}[b]{0.32\textwidth}
        \includegraphics[width=\textwidth]{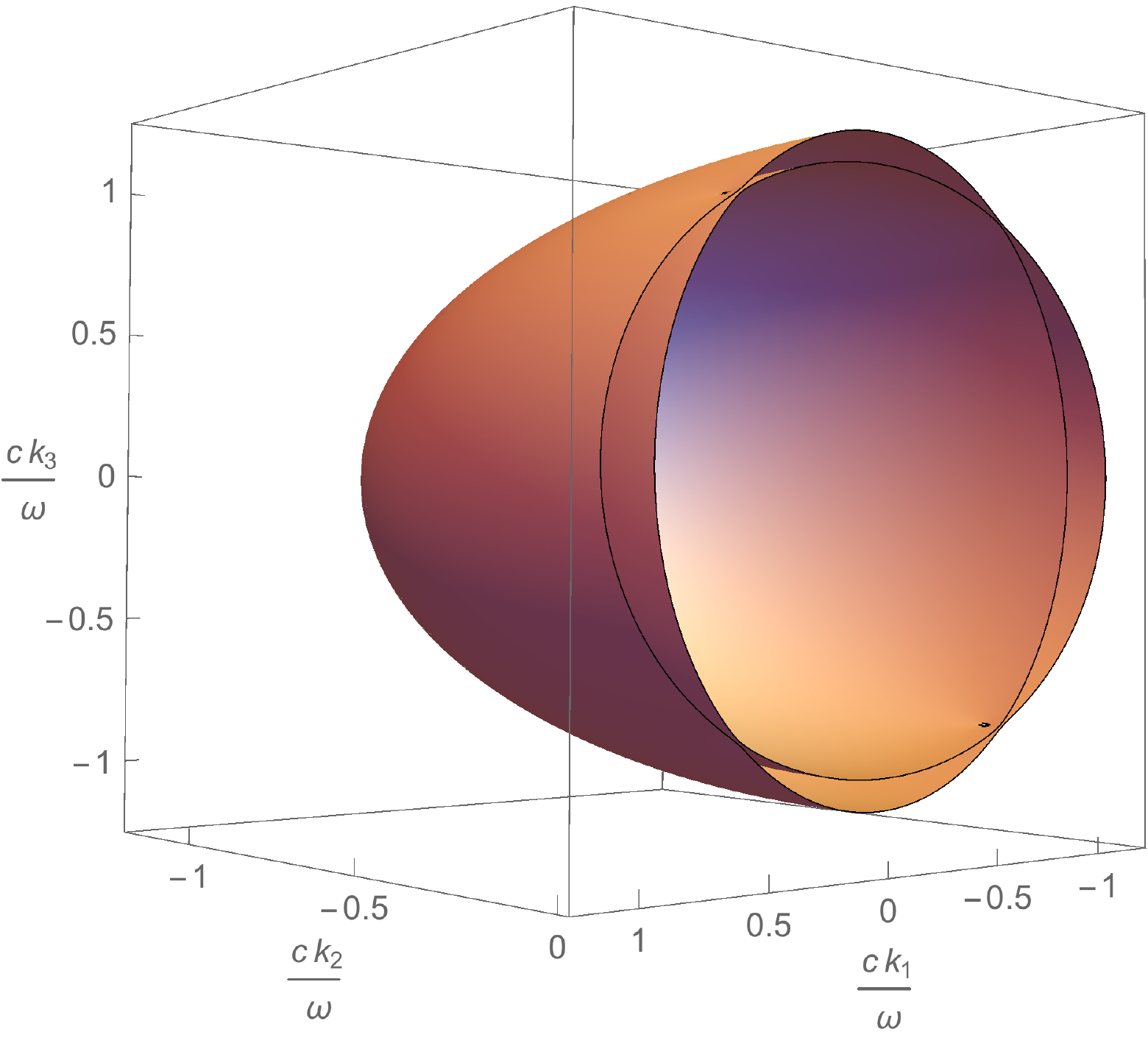}
        \caption{$\delta=10$}
    \end{subfigure}
    ~ 
    \begin{subfigure}[b]{0.32\textwidth}
        \includegraphics[width=\textwidth]{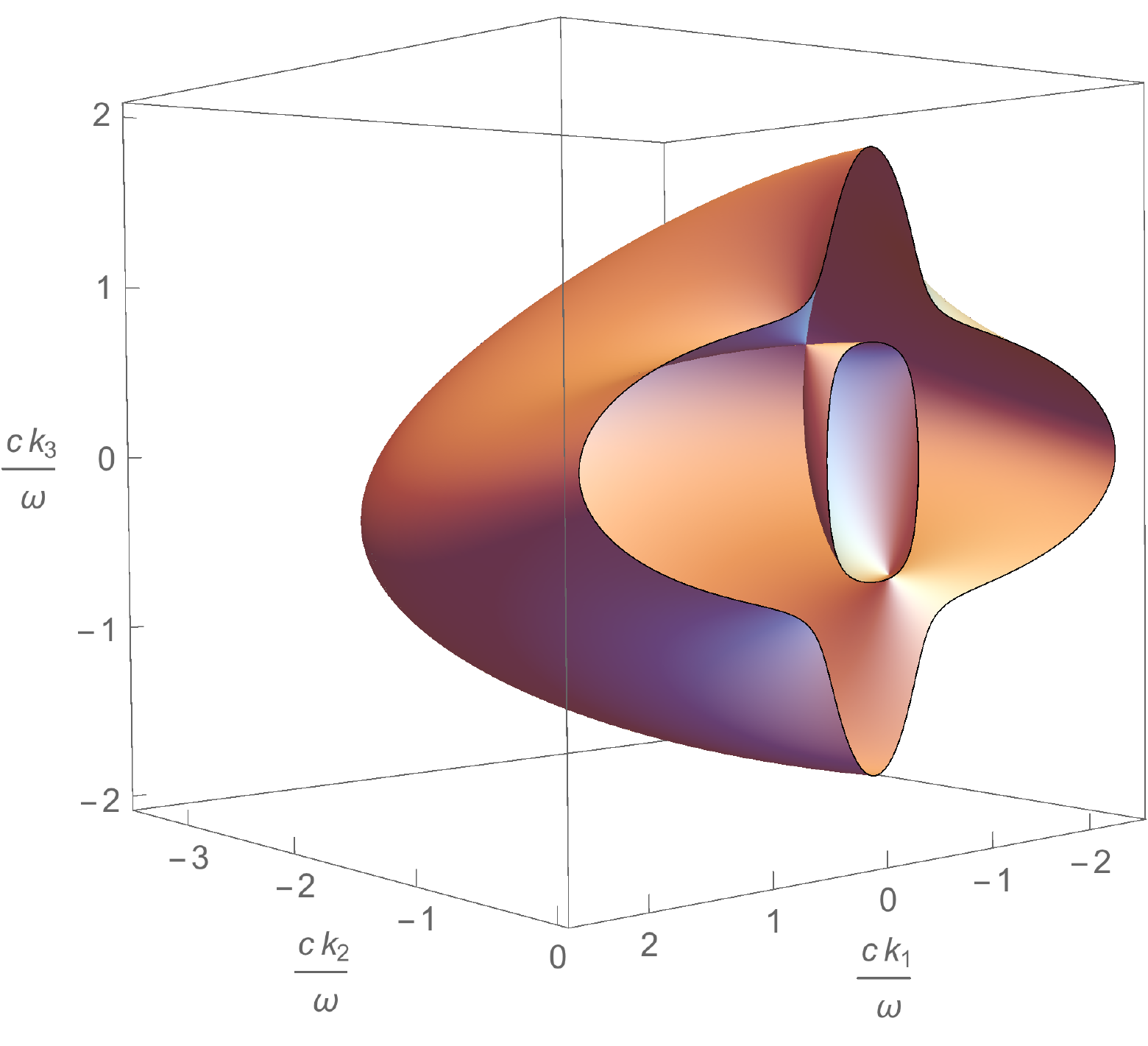}
        \caption{$\delta=1$}
    \end{subfigure}\vspace{10pt}
   
    \begin{subfigure}[b]{0.32\textwidth}
        \includegraphics[width=\textwidth]{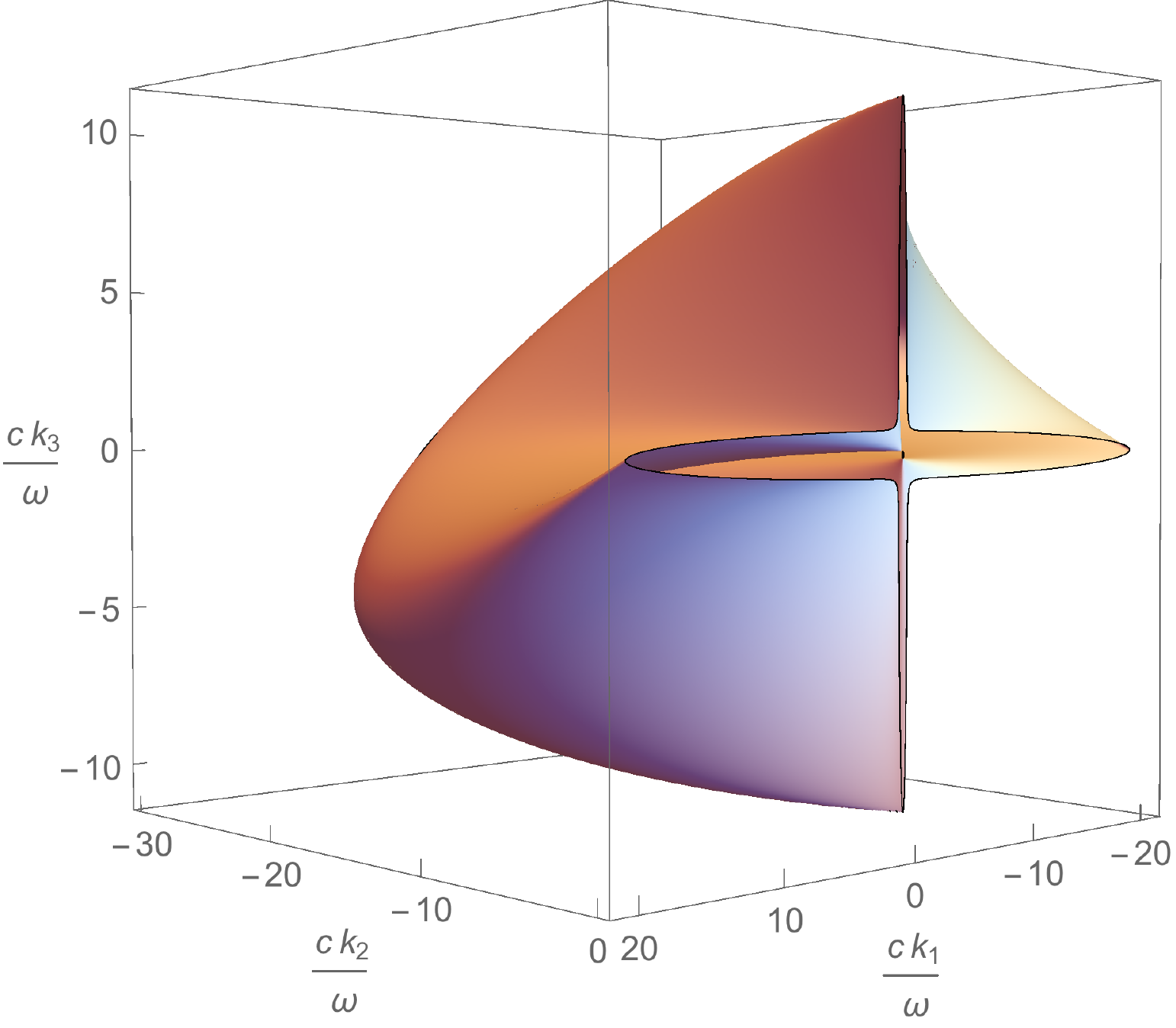}
        \caption{$\delta=0.1$}
    \end{subfigure}
    ~ 
   \begin{subfigure}[b]{0.32\textwidth}
        \includegraphics[width=\textwidth]{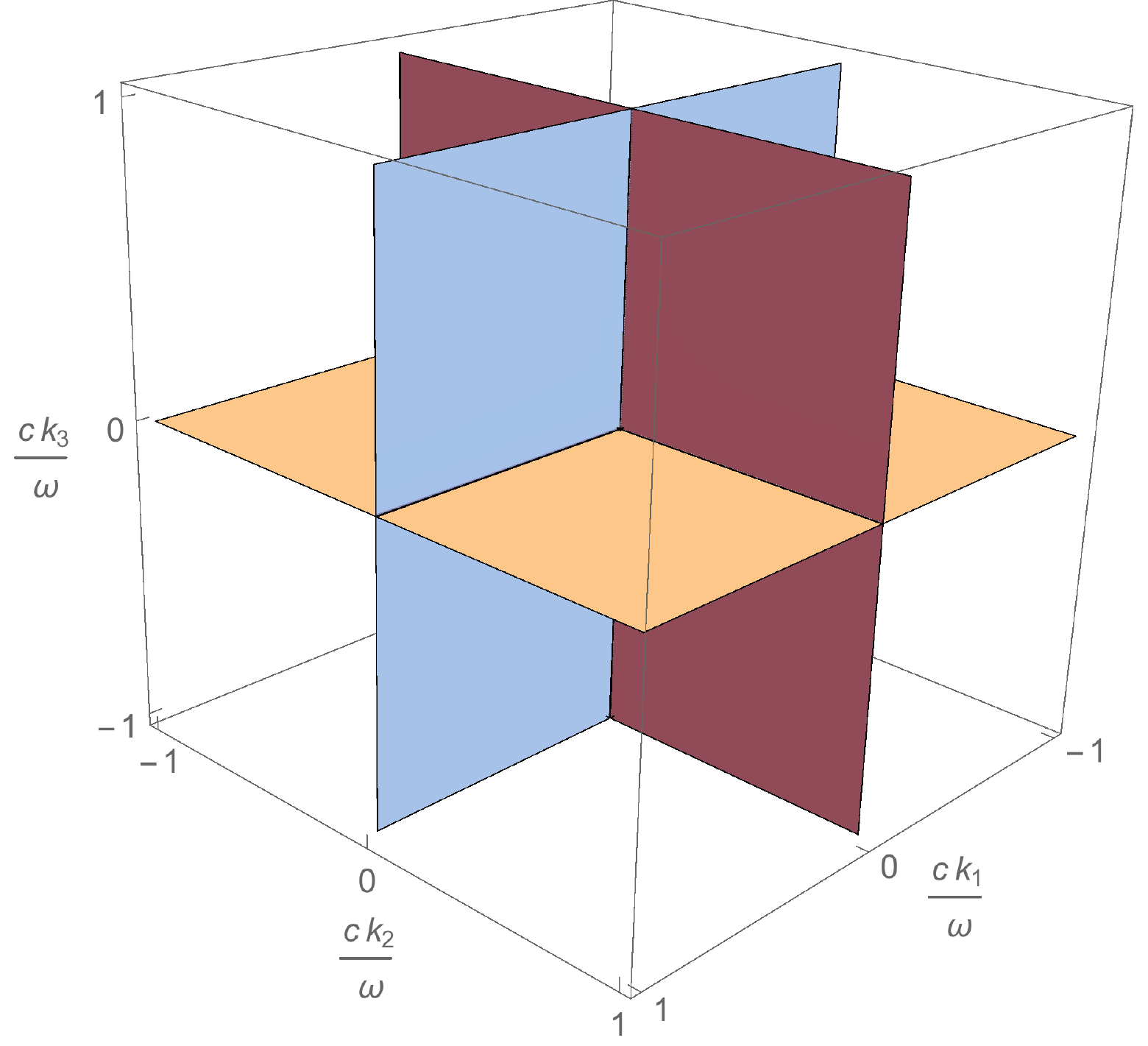}
        \caption{$\delta=0$}
    \end{subfigure}
    \caption{Dispersion surfaces for a magnetoelectric medium defined by \r{EEME} and \r{hoB}. For $\de\ra0$ the medium becomes an extreme magnetoelectric (EME) medium whose dispersion surface consists of three orthogonal planes extending to infinity. For finite values of $\de$, cross sections of the quartic surfaces are shown.}
\label{Fig2}
\end{figure}

\begin{itemize}
\item For the skewon special case with $\=\A=\=\B{}^T$, we have that $f(\=\A,\=\A)=0$, whence \r{dispAB} is again identically satisfied for any $\#k$. 
\item For the principal-axion special case with $\A_i=-\B_i$, \r{dispAB} reduces to the simpler form
\e k_1k_2k_3(\A_1-\A_2)(\A_2-\A_3)(\A_3-\A_1)=0. \l{k123}\f
\item When the dyadics $\=\A$ and $\=\B$ are related so that the function $f(\=\A,\=\B{}^T)$  vanishes, the dispersion equation is satisfied for any $\#k$ and the medium becomes an example of an NDE medium. For the principal-axion case \r{k123}, this happens when at least two of the eigenvalues $\A_i$ coincide. 
\item For $f(\=\A,\=\B{}^T)\not=0$, the dispersion equation \r{dispAB} is reduced to $k_1k_2k_3=0$, whence at least one of the components $k_i$ must vanish. This means that any wave vector $\#k\not=0$ must be parallel to one of the three eigenplanes defined by the three pairs of eigenvectors $\{\#a_i,\#a_j\}$, while $\#k=0$ for all directions outside these three eigenplanes. Thus, instead of a well-defined dispersion surface, the possible $\#k$ vectors specify three planes intersecting along the three eigenvectors $\#a_i$, with a common point at the origin. This is demonstrated in Figure 1a, which presents the dispersion surface as truncated by two spherical surfaces. For comparison, in Figure 1b, the resulting dispersion surface is shown for two dyadics $\=\A$ and $\=\B$ created by two randomly chosen matrices possessing real eigenvectors. 
\end{itemize}

The emergence of the unusual dispersion surface consisting of three planes can be visualized by adding permittivity and permeability dyadics to the medium equations as
\e \am \#D \\ \#H\a = \amm \=\A & \de\E_o\=I \\ \de\=I/\M_o & \=\B\a\.\am \#B\\ \#E\a , \l{EEME}\f
and letting $\de$ approach zero. If the principal-axion EME medium is defined by the normalized symmetric dyadics
\e \h_o\=\B= -\h_o\=\A= \#u_1\#u_1 + 2\#u_2\#u_2 + 4\#u_3\#u_3, \l{hoB}\f
where the normalized eigenvectors $\#u_i$ are orthogonal, the wave vectors are restricted to the orthogonal eigenplanes $\#u_i\.\#r=0$ for $i=1,2$ and $3$. 

The effect of $\de$ is portrayed in Figure \ref{Fig2}, showing cross cuts of the corresponding quartic dispersion surfaces by a plane parallel to one of the eigenplanes. It is seen that, starting from a spherical dispersion surface for $\de=\oo$ (not shown in the figure), for decreasing values of $\de$ it is split in two surfaces, one of which shrinks towards the origin ($\#k\ra0$), while the other approaches the planes corresponding to the three orthogonal eigenplanes.  Then, any $\#k$ vector on the three planes is a valid solution. Dispersion equations for less extreme magnetoelectric media can also yield exotic dispersion surfaces, as has been demonstrated in \cite{AlbertoHehl16}.

The dispersion equation of a generic linear medium has the form $D(\omega,\#k)=0$, where the left-hand side is a homogeneous quartic polynomial in $\omega$ and $\#k$. Linear media can be classified according to how, if at all, $D(\omega,\#k)$ factorizes in homogeneous polynomials of lower degree. In the general case \cite{Hehl,MDEM}, and in complicated media such as biaxial crystals \cite{Jenkins,Born}, no factorization takes place. Media such that $D(\omega,\#k)$ is the product of two quadratic polynomials are well studied. Examples include hyperbolic metamaterials \cite{Poddubny,Ferrari}, uniaxial media \cite{Jenkins,Born}, nematic liquid crystals \cite{Obukhov2012} and decomposable (DCM) media \cite{LindellDCM,Dahl2013}. If the two quadratic factors coincide, the material displays no birefringence. Isotropic media belong to this class and, to a great extent, summarize its properties \cite{Favaro2011,Dahl2012}. The present article considers two seemingly unexplored factorisations of $D(\omega,\#k)$. For generic EME media, the quartic dispersion polynomial is the product of a linear polynomial, $\omega$, and a cubic irreducible one, $\det\hspace{0.5pt}\=D(\#k)$. When $\=\A$ and $\=\B{}^T$ satisfy \r{Commute}, this latter quantity becomes factorable, see \r{dispAB}. Thereby, EME media with commuting dyadics are examples of materials such that $D(\omega,\#k)$ is the product of four linear polynomials in $\omega$ and $\#k$.

\subsection{Fields}

In an EME medium with commuting dyadics, see \r{Commute}, the components of the electric field satisfy the following relations extracted from \r{DEAB},
\ea (\A_1+\B_2)k_3E_2 &=& (\A_1+\B_3)k_2E_3 ,  \l{AB13}\\
(\A_2+\B_3)k_1E_3  &=& (\A_2+\B_1)k_3E_1 , \l{AB21}\\
(\A_3+\B_1)k_2E_1 &=& (\A_3+\B_2)k_1E_2 . \l{AB32}\fa
Equating the product of the left-hand sides with that of the right-hand sides yields the dispersion equation \r{dispAB}, multiplied by $E_1E_2E_3$. 

To understand how the $\#k$ vector and the field vectors are related, let us assume that the dyadics $\=\A$ and $\=\B$ obey $f(\=\A,\=\B{}^T)\not=0$, see \r{fAB}. In this scenario, the dispersion equation requires that $k_i=0$ for at least one value of $i=1,2,3$. Geometrically, the dispersion surface is made up of three eigenplanes, and wave vectors in all other directions must vanish, $\#k\!=\!0$. Let us examine the case $k_{3}=0$ with $\#k\neq 0$, so that \r{AB13} and \r{AB21} yield $E_{3}=0$. Similar results are obtained for the other two cases, $k_{1}=0$ and $k_{2}=0$, by a change of indices. To derive the electromagnetic fields, it is convenient to express the wave vector in the relevant eigenplane as
\e \#k = k(\#a_1\cos\VF + \#a_2\sin\VF). \f
By using \r{AB32}, the Maxwell equation \r{kxE}, and the medium law \r{ME} with \r{ABsum}, one then calculates that
\ea \#E&=& \#a_{1}S(\A_3+\B_2)\cos\VF + \#a_{2}S(\A_3+\B_1)\sin\VF,\l{vFe}\\
\o\#B &=& \#a_3'kS(\B_1-\B_2)\sin\VF\cos\VF, \l{vFb}\\
\o\#D &=& \#a_3'kS\A_3(\B_1-\B_2)\sin\VF\cos\VF,\l{vFd}\\
\#H &=& \#a_{1}S\B_1(\A_3+\B_2)\cos\VF + \#a_{2}S\B_2(\A_3+\B_1)\sin\VF,\l{vFh}\fa
where $S$ is an arbitrary scalar. It is observed that the angle $\VF$ rotates the fields $\#E$ and $\#H$ in the plane of $\{\#a_1,\#a_2\}$ while affecting the magnitude of the vectors $\#B$ and $\#D$ normal to that plane. From
\e \#E\x\#H = \#a_3' (S^2/2)(\B_1-\B_2)(\A_3+\B_2)(\A_3+\B_1)\sin(2\VF), \f
we can further see that, if $\VF=n\pi/2$, the vectors $\#E,\#H$ and $\#k$ are parallel to each other and to either $\#a_1$ ($n$ even) or $\#a_2$ ($n$ odd). These cases correspond to  waves propagating along an eigendirection of the dyadics $\=\A$ and $\=\B$ with $\#B=\#D=0$. 

Substituting $\B_i=-\A_i$ into \r{vFe} -- \r{vFh} determines the field vectors in principal-axion EME media with commuting dyadics. One must however impose that the three eigenvalues $\A_i$ are distinct, otherwise $f(\=\A,\=\B{}^{T})$ vanishes, and the above expressions fail.

For the same reason, \r{vFe} -- \r{vFh} are not valid for pure-skewon EME media with commuting dyadics. In this case, the field vectors are given by \r{Esk} -- \r{DSk} together with the expansion of $\=\A$ in \r{ABsum}.

If $\=\A$ and $\=\B{}^{T}$ satisfy \r{Commute}, and have identical eigenvalues, we obtain a pure-axion medium (PEMC). Then, \r{kxE} and \r{kxH} coincide, which leaves a lot of freedom for the fields \cite{Jancewicz}.

\section{Uniaxial EME medium}

As an example of EME media with vanishing function $f(\=\A,\=\B{}^T)$ of \r{fAB}, let us consider uniaxial media with symmetric commuting dyadics $\=\A$ and $\=\B$ defined by 
\e \=\A = \A_t\=I_t+ \A_3\#u_3\#u_3, \ \ \ \=\B = \B_t\=I_t+ \B_3\#u_3\#u_3, \ \ \ \ \=I_t=\#u_1\#u_1+ \#u_2\#u_2,\l{uniax}\f
where the vectors $\#u_i$ make an orthonormal vector basis. Here, $()_t$ denotes component transverse to the axial direction $\#u_3$. In the general case, none of the Hehl--Obukhov components $\=\A_i$ defined by \r{A1} -- \r{A3} vanish. The number of free parameters for this medium is reduced to $2 (\#u_3)  + 4 (\A_t,\A_3,\B_t,\B_3) = 6$.

\subsection{Plane wave in uniaxial medium}

Since the uniaxial EME medium makes another example of an NDE medium, there is no a priori restriction for the choice of the wave vector $\#k$. From \r{DkE} and \r{Dk} we have
\e \#u_3\.\=D(\#k)\.\#E =0,\ \ \ \Ra\ \ \ (\A_3+\B_t)\#k_t\x\#E_t=0, \f
whence, assuming $\A_3+\B_t\not=0$, the fields transverse to the axial direction can be represented by
\e \#E_t = U'\#k_t,\ \ \ \#H_t=\B_t U'\#k_t,\ \ \ \#k_t=\=I_t\.\#k, \l{uniU}\f
for some factor $U'$. Because of
\e \o\#u_3\.\#B = \#u_3\.\#k_t\x\#E_t =0,\ \ \ \o\#u_3\.\#D = -\#u_3\.\#k_t\x\#H_t=0, \f
the axial components of $\#B$ and $\#D$ vectors actually vanish in the uniaxial EME medium,
\e \#u_3\.\#B=\#u_3\.\#D=0.\l{uniB3}\f
Expanding
\ea (\#k\x\#E-\o\#B)_t &=& k_3\#u_3\x\#E_t-E_3\#u_3\x\#k_t -\o\#B_t=0, \\
 (\#k\x\#H+\o\#D)_t &=& k_3\B_t\#u_3\x\#E_t - \B_3 E_3\#u_3\x\#k_t + \o\A_t\#B_t =0, \fa
inserting \r{uniU} and eliminating $\#u_3\x\#k_t$, we have either $\#k_t=0$ and $\#B_t=0$ or
\e E_3=(\A_t+\B_t)k_3U,\ \ \ \ U=U'/(\A_t+\B_3). \l{uniUE3}\f
Since the first possibility leads to vanishing of all fields in the medium, we can omit that possibility. Applying \r{uniUE3}, the fields can be constructed as
\ea \#E &=& U((\A_t+\B_3)\#k_t +(\A_t+\B_t)k_3\#u_3), \l{EE}\\
\#H &=& U(\B_t(\A_t+\B_3)\#k_t +\B_3(\A_t+\B_t)k_3\#u_3)= \=\B\.\#E, \l{HH}\\
\o\#B &=& U(\B_3-\B_t)k_3\#u_3\x\#k_t, \l{BB}\\
\o\#D &=& U\A_t(\B_3-\B_t)k_3\#u_3\x\#k_t=\A_t\o\#B. \l{DD}\fa
It is remarkable that, in a given uniaxial EME medium, the $\#k$ vector can be freely chosen, after which the field polarizations are uniquely determined.

\subsection{Interface of uniaxial EME medium}

Assuming continuity of fields through a planar interface $\#n\.\#r=0$  of an isotropic medium and the uniaxial EME medium defined by \r{uniax}, the interface conditions for the sum of the incident and reflected fields from \r{EE} and \r{HH} become
\ea \#n\x(\#E^i+ \#E^r) &=& U\#n\x((\A_t+\B_3)\#k_t +(\A_t+\B_t)k_3\#u_3) \l{nE}\\
\#n\x(\#H^i+\#H^r) &=& U\#n\x(\B_t(\A_t+\B_3)\#k_t +\B_3(\A_t+\B_t)k_3\#u_3) \l{nH}\\
\o\#n\.(\#B^i+\#B^r) &=& U(\B_3-\B_t)k_3\#n\.\#u_3\x\#k_t, \l{nBB}\\
\o\#n\.(\#D^i+\#D^r) &=& U\A_t(\B_3-\B_t)k_3\#n\.\#u_3\x\#k_t, \l{nDD}\fa
where $\#n$ denotes the unit vector normal to the interface. In certain cases there may arise induced surface sources at the interface to balance the fields on both sides.
 
As an example, let us consider reflection from a planar interface with normal of the interface coinciding with the axis of the medium, $\#n=\#u_3$. Because of rotational symmetry, we can assume for simplicity that incident and reflected wave vectors are of the form
\e \#k^i = \#u_1 k_1 + \#u_3 k_3^i, \ \ \ \#k^r = \#u_1k_1 - \#u_3 k_3^i. \f
The wave vector tangential to the interface $\#k_t=\#u_1k_1\not=0$ is shared by the plane wave transmitted into the EME medium. The conditions for the incident and reflected fields \r{nE}, \r{nH}, \r{nBB} and \r{nDD} can be written as 
\ea (\#E^i+ \#E^r)_t &=& U(\A_t+\B_3)\#u_1 k_1, \l{nE1}\\
(\#H^i+\#H^r)_t &=& U\B_t(\A_t+\B_3)\#u_1 k_1, \l{nH1}\\
\o\#u_3\.(\#B^i+\#B^r) &=& 0, \l{nBB1}\\
\o\#u_3\.(\#D^i+\#D^r) &=& 0. \l{nDD1}\fa
From \r{nBB1} and \r{nDD1} it appears that the interface acts as a surface with DB boundary conditions \cite{DB,AP09}. However, because of the unknown quantity $U$ in \r{nE1} and \r{nH1}, we must verify this assumption.

It is known that the DB boundary has eigenfields polarized TE and TM with respect to the normal direction, and for the eigenfields, the DB boundary is equivalent to PEC and PMC boundaries, respectively \cite{DB,AP09,AP10}. Let us verify this by considering  reflection of incident TE and TM waves from the uniaxial EME interface. The two waves are defined by the conditions 
\ea \#u_3\.\#E_{TE}^i&=&0, \l{ETE}\\
 \#u_3\.\#H_{TM}^i&=&0. \l{HTM} \fa

For the TE wave, from \r{nDD1} we obtain
\e \#u_3\.\#E{}_{TE}^r=0. \f
Applying the orthogonality conditions
\e \#k^i\.\#E^i = (\#u_1k_1 + \#u_3k_3^i)\.(\#u_1E_{TE1}^i+ \#u_2E_{TE2}^i)= k_1E_{TE1}^i=0, \f 
\e \#k^r\.\#E^r = (\#u_1k_1 - \#u_3k_3^i)\.(\#u_1E_{TE1}^r+ \#u_2E_{TE2}^r)= k_1E_{TE1}^r=0, \f 
we have
\e E_{TE1}^i= E_{TE1}^r=0. \f
From \r{nE1} we obtain
\ea  \#u_2\.(\#E_{TE}^i+ \#E_{TE}^r)_t &=& 0, \\
\#u_1\.(\#E_{TE}^i+ \#E_{TE}^r)_t &=& U(\A_t+\B_3)k_1=0.\fa
The latter yields
\e  U=0, \f
whence the fields in the EME medium must actually vanish. The reflected field satisfies
\e \#E_{TE}^r = -\#E_{TE}^i = -\#u_2 E_{TE2}^i, \l{EPEC}\f
which can be recognized as the PEC boundary condition for the electric field. Expanding
\ea \o\M_o(\#H_{TE}^i+ \#H_{TE}^r)_t &=& (\#k^i\x\#E_{TE}^i+ \#k^r\x\#E_{TE}^r)_t \nonumber\\
&=& -\#u_1k_3^i(E_{TE2}^i-E_{TE2}^r) \nonumber\\
&=& -2\#u_1k_3^iE_{TE2}^i\not=0, \fa
we find that \r{nH1} cannot be valid for $U=0$. Thus, there must exist an induced surface current $\#J_s$, whence \r{nH1} must be replaced by \cite{Methods}
\e (\#H^i+\#H^r)_t = U\B_t(\A_t+\B_3)\#k_t -\#u_3\x\#J_s = -\#u_3\x\#J_s. \f
The surface current can be found from
\e \#J_s = \frac{1}{\o\M_o}\#u_3\x(-2\#u_1k_3^iE_{TE2}^i) = -\frac{2k_3^i}{\o\M_o}\#u_2E_{TE2}^i . \f
\r{nDD1} must also be upgraded as
\e \o\#u_3\.(\#D^i+\#D^r) = \#k_t\.\#J_s = -\frac{2k_3^i}{\o\M_o}k_1\#u_1\.\#u_2E_{TE2}^i=0,\f
but this does not actually change \r{nDD1}. The magnetic fields are found as
\ea \#H_{TE}^i &=& \frac{1}{\o\M_o}\#k^i\x\#u_2 E_{TE2}^i = \frac{1}{\o\M_o}(k_1\#u_3 -k_3^i\#u_1)E_{TE2}^i \\ 
\#H_{TE}^r &=& \frac{-1}{\o\M_o}\#k^r\x\#u_2 E_{TE2}^i = -\frac{1}{\o\M_o}(k_1\#u_3 +k_3^i\#u_1)E_{TE2}^i \fa
whence we can verify the relations
\e \#u_3\.(\#B_{TE}^i+\#B_{TE}^r)= \o\M_o\#u_3\.(\#H_{TE}^i+\#H_{TE}^r)=0,\f
and
\e (\#H_{TE}^i+\#H_{TE}^r)_t = -\frac{2k_3^i}{\o\M_o}\#u_1E_{TE2}^i = -\#u_3\x\#J_s. \f
Thus, the interface of the uniaxial EME medium acts as a PEC boundary for the TE plane wave. 

For the TM case we can proceed similarly and find that the interface acts as a PMC boundary. Changing symbols, we can rewrite the above results as
\e U=0 \f
and
\e \#H_{TM}^r = -\#H_{TM}^i= -\#u_2 H_{TM2}^i.\f
In this case we must change the conditions \r{nE1} and \r{nBB1} to
\ea (\#E^i+ \#E^r)_t &=& U(\A_t+\B_3)\#u_1 k_1 +\#u_3\x\#J_{ms} = \#u_3\x\#J_{ms},\\
\o\#u_3\.(\#B^i+\#B^r) &=& -\#k_t\.\#J_{ms}. \fa
The magnetic surface current becomes
\e \#J_{ms} =  -\frac{2k_3^i}{\o\E_o}\#u_2H_{TM2}^i ,\ \ \ \ \#k_t\.\#J_{ms}=0. \f
and
\e \#E_{TM}^r = \frac{-1}{\o\E_o}(k_1\#u_3 +k_3^i\#u_1)H_{TM2}^i. \f
To summarize, the interface of the uniaxial EME medium acts as a DB boundary because the interface conditions \r{nBB1} and \r{nDD1} do not contain surface sources.

\section{Conclusion}
A novel class of electromagnetic media, labeled as that of extreme magnetoelectric (EME) media, was introduced in terms of two medium dyadics, $\=\A$, creating electric polarization through the magnetic field and $\=\B$, creating magnetic polarization through the electric field. For the more general bi-anisotropic media, $\=\A$ and $\=\B$ are known as magnetoelectric dyadics, but, in this extreme case, the permittivity and inverse permeability dyadics are assumed to vanish. Plane wave propagation in EME media was considered with several examples of special cases. It was shown that, for a fixed nonzero frequency $\omega$, the dispersion equation corresponding to an EME medium is cubic and homogeneous in the $\#k$ vector, whence the magnitude of $\#k$ is not restricted. Actually, the dispersion surface is a conical surface defined by the origin and a closed curve, or a set of closed curves, on the unit sphere. However, applying 4D formalism (not considered here), one can show that the dispersion equation is actually quartic with $\o=0$ as another possible solution. In many special cases the dispersion equation is reduced to an identity, satisfied for any $\#k$ vector, whence the medium is an example of NDE (no dispersion equation) medium. We demonstrated that, for non-NDE EME media with commuting dyadics $\=\A$ and $\=\B{}^T$, the dispersion surface is reduced to three intersecting eigenplanes, defined by the three pairs of common eigenvectors of the two dyadics. Wave reflection from an interface of a uniaxial EME medium was considered when the axis of the medium is normal to the interface, in which case the interface was shown to act as a DB boundary with vanishing normal components of the $\#D$ and $\#B$ vectors. To extend the present analysis, complex solutions to the dispersion equation and the corresponding field equations, ignored here, should be studied.

\section*{Acknowledgments}
The authors thank Professor Friedrich W.\ Hehl for many comments on the draft of this paper. A.F.\ gratefully acknowledges financial support from the Gordon and Betty Moore Foundation.

\section*{Appendix 1: Hehl--Obukhov decomposition}
The set of medium dyadics of any electromagnetic medium can be most naturally decomposed in three subsets,
\e \amm \=\A & \=\E'\\ \=\M{}^{-1} & \=\B\a =  \amm \=\A_1 & \=\E'_1\\ \=\M{}^{-1}_1 & \=\B_1\a + \amm \=\A_2 & \=\E'_2\\ \=\M{}^{-1}_2 & \=\B_2\a + \amm \=\A_3 & \=\E'_3\\ \=\M{}^{-1}_3 & \=\B_3\a , \f 
respectively labeled as the principal, skewon and axion parts of the medium by Hehl and Obukhov \cite{Hehl}. Such a decomposition can be most logically introduced by applying the four-dimensional formalism, details of which can be found in \cite{Hehl,MDEM}. For the 3D Gibbsian dyadic quantities considered here, the decomposition can be defined in somewhat awkward manner by reformulating the medium equations \r{DBE} and \r{HBE} as
\e \am \#D\\ \#H\a = \amm -\=\E' & \=\A\\ -\=\B & \=\M{}^{-1} \a \.\am -\#E\\ \#B\a . \f
The Hehl--Obukhov decomposition can now be expressed as \cite{Hehl}
\e   \amm -\=\E' & \=\A\\ -\=\B & \=\M{}^{-1} \a = \amm -\=\E'_1 & \=\A_1\\ -\=\B_1 & \=\M{}_1^{-1} \a + \amm -\=\E'_2 & \=\A_2\\ -\=\B_2 & \=\M{}_2^{-1} \a+ \A_3\amm 0 & \=I\\ \=I & 0 \a, \f
where the principal part and the skewon part are respectively represented by symmetric and antisymmetric dyadic matrices. However, the symmetric axion part must be {extracted} from the first symmetric matrix through the condition $\tr\=\A_1=\tr\=\B_1=0$. Thus, the decomposed medium dyadics satisfy the following relations:
\begin{enumerate}
\item Principal part: $\=\E'_1=\=\E'{}_1^T$,\ \ \ $\=\M{}_1^{-1}=\=\M{}_1^{-1T}$,\ \ \  $\=\A_1=-\=\B{}_1^T$,\ \ \   $\tr\=\A_1=0${;}
\item Skewon part: $\=\E'_2=-\=\E'{}_2^T$,\ \ \  $\=\M{}_2^{-1}=-\=\M{}_2^{-1T}$,\ \ \  $\=\A_2=\=\B{}_2^T${;} 
\item Axion part: $\=\E'_3=0,\ \ \  \=\M{}^{-1}_3=0$,\ \ \  $\=\A_3=-\=\B_3=\A_3\=I${.}
\end{enumerate}

\end{document}